\documentclass[onecolumn,showpacs]{revtex4}

\usepackage{graphicx}
\usepackage{dcolumn}
\usepackage{bm}

\input epsf

\begin{document}

\title{\Large Generalized Ricci dark energy in Horava-Lifshitz gravity}

\author{Surajit Chattopadhyay}
\email{surajit_2008@yahoo.co.in,
surajit.chattopadhyay@rediffmail.com} \affiliation{Department of
Computer Application (Mathematics Section), Pailan College of
Management and Technology, Bengal Pailan Park, Kolkata-700 104,
India.}

\date{\today}

\begin{abstract}

 In this
letter, we have considered generalized Ricci dark energy in the
Horava-Lifshitz gravity. We have reconstructed the Hubble's
parameter in terms of fractional densities. We have viewed the
equation of state parameter in this situation. Also, we have
examined the behavior of deceleration parameter and investigated
the nature of the statefinder diagnostics. The equation of state
parameter has exhibited quintessence-like behavior and from the
plot of the deceleration parameter we have observed an ever
accelerating universe.
\end{abstract}

\maketitle

 The accelerating cosmic expansion of the present universe has been strongly
 confirmed by the cosmic microwave
background radiation (CMBR) \cite{Spergel} and Sloan Digital Sky
Survey (SDSS) \cite{Adelman}. An exotic form of negative pressure
matter called dark energy (DE) is used to explain this
acceleration. Extensive reviews on DE are available in
\cite{Copeland}, \cite{Padmanabhan1}, \cite{Li} and \cite{Cai}.
The most powerful quantity of DE is its equation-of-state (EoS)
effectively defined as $w_{DE}=p_{DE}/\rho_{DE}$, where $p_{DE}$
and $\rho_{DE}$ are the pressure and energy density respectively.
All DE models can be classified by the behaviors of equations of
state as following \cite{Cai}:
\begin{itemize}
    \item Cosmological constant: its EoS is exactly equal to $-1$, that is $w_{DE}=-1$.
    \item Quintessence: its EoS remains above the cosmological constant boundary, that is $w_{DE}\geq-1$.
    \item Phantom: its EoS lies below the cosmological constant boundary, that is $w_{DE}\leq-1$.
    \item Quintom: its EoS is able to evolve across the cosmological constant boundary.
\end{itemize}
An approach to the problem of DE arises from holographic principle
that states that the number of degrees of freedom related directly
to entropy scales with the enclosing area of the system
\cite{Setare1}. Works on \emph{holographic dark energy} (HDE)
include the references \cite{Zhang}, \cite{LiHDE},
\cite{Zhanghtachyon} and \cite{Wu}. There are many papers about
holographic reconstruction of different dark energy models.
References in this regard are given in \cite{holographic}.
Inspired by the HDE models, Gao et al \cite{Gao} proposed a DE
model dubbed as ``Ricci dark energy" (RDE), where the energy
density of
 the universe is proportional to the Ricci scalar
 $R=-6\left(\dot{H}+2H^{2}+\frac{k}{a^{2}}\right)$, where dot denotes a
derivative with respect to time $t$ and $k$
 is the spatial curvature. The energy density of RDE is \cite{Gao}

 \begin{equation}
\rho_{R}=3 c^{2}\left(\dot{H}+2H^{2}+\frac{k}{a^{2}}\right)
\end{equation}

where $c$ is a dimensionless parameter which will determine the
evolution behavior of RDE. When $c^{2}<1/2$, the RDE will exhibit
a quintomlike behavior. With the proposed form of energy density,
the energy density can be solved from the Friedmann equation as

\begin{equation}
\rho_{R}=\frac{\alpha}{2-\alpha}\Omega_{m0}e^{-3\ln
a}+f_{0}e^{-\left(4-\frac{\alpha}{2}\right)\ln a}
\end{equation}

where, $\alpha=\frac{8\pi c^{2}}{3}$. Thus the RDE has one part
which evolves like nonrelativistic matter $(\sim e^{-3\ln a})$ and
another part which is slowly increasing with decreasing redshift
\cite{Gao}. Subsequently, the pressure of RDE can be obtained as
\cite{Gao}

\begin{equation}
p_{R}=-\left(\frac{2}{3\alpha}-\frac{1}{3}\right)f_{0}e^{-\left(4-\frac{\alpha}{2}\right)\ln
a}
\end{equation}

In the studies subsequent to the reference \cite{Gao}, various
aspects of RDE have been investigated. In the said reference
\cite{Gao}, it was shown that this model can avoid the causality
problem and naturally solve the coincidence problem of dark
energy. Xu et al \cite{Xu} revisited the RDE model and used cosmic
observations to constrain the model parameters. Feng and Li
\cite{FengLi} investigated the viscous RDE model by assuming that
there is bulk viscosity in the linear barotropic fluid and the
RDE. Feng \cite{Feng} reconstructed $f(R)$ theory from RDE. A
correspondence between various dark energy candidates and RDE was
studied by Chattopadhyay and Debnath \cite{ChattoRDE}. A
generalized model has been designed by Xu et al \cite{Xu1} to
included HDE and RDE by introducing a new parameter which balances
holographic and Ricci dark energy model. Reference \cite{Xu1}
considered a generalized versions of holographic and Ricci dark
energy and taken the energy densities for generalized holographic
dark energy (GHDE) and generalized Ricci dark energy (GRDE) as

\begin{equation}
\rho_{GH}=3c^{2}M_{pl}^{2}~f\left(\frac{R}{H^{2}}\right)H^{2}
\end{equation}

\begin{equation}
\rho_{GR}=3c^{2}M_{pl}^{2}~g\left(\frac{H^{2}}{R}\right)R
\end{equation}

where $f(x)$ and $g(y)$ are positive defined functions of the
dimensionless variables $x=R/H^{2}$ and $y=H^{2}/R$ respectively.
Holographic and Ricci dark energy models are recovered when the
function $f(x)=g(y)\equiv1$. Also, when the function $f(x)=x$ and
$g(y)=y$, the holographic and Ricci dark energy exchange each
other. The functions can be written as \cite{Xu1}

\begin{equation}
f\left(\frac{R}{H^{2}}\right)=1-\epsilon
\left(1-\frac{R}{H^{2}}\right)
\end{equation}

\begin{equation}
g\left(\frac{H^{2}}{R}\right)=1-\eta
\left(1-\frac{H^{2}}{R}\right)
\end{equation}

where $\epsilon$ and $\eta$ are parameters. When $\epsilon=0 (\eta
=1)$ or $\epsilon=1 (\eta =0)$, the generalized energy density
becomes the holographic (Ricci) and Ricci (holographic) dark
energy density respectively. If $\epsilon=1-\eta$, then GHDE and
GRDE are equivalent. Endeavor of the present work is to discuss
the GRDE in Horava-Lifshitz (HL) gravity \cite{Horava},
\cite{Calcagni}, \cite{Kiritsis}. The basic idea of the HL
approach is to modify the UV behavior of the general theory so
that the theory is perturbatively renormalizable \cite{Kluson}. In
this theory the local Lorentz invariance is abandoned, but it is
restored as an approximate symmetry at low energies
\cite{Carloni}. Review on HL gravity is available in the reference
\cite{Kiritsis}.

\subsection{An overview of Horava-Lifshitz gravity}

We briefly review the scenario where the cosmological evolution is
governed by HL gravity. The dynamical variables are the lapse and
shift functions, $N$ and $N_{i}$ respectively, and the spatial
metric $g_{ij}$. In terms of these fields the full metric is
written as \cite{Jamil}, \cite{Colgain}

\begin{equation}
ds^{2}=-N^{2}dt^{2}+g_{ij}(dx^{i}+N^{i}dt)(dx^{j}+N^{j}dt)
\end{equation}

where indices are raised and lowered using $g_{ij}$. The scaling
transformation of the coordinates reads: $t\rightarrow l^{3}t$ and
$x^{i}\rightarrow lx^{i}$.

The action of the HL gravity is given by \cite{Jamil},
\cite{Colgain}

\begin{equation}
\begin{array}{c}
  I=dt
  \int
  dtd^{3}x(\mathcal{L}_{0}+\mathcal{L}_{1}+\mathcal{L}_{m})\\\\
  \mathcal{L}_{0}=\sqrt{g}N \left[\frac{2}{\kappa^{2}}(K_{ij}K^{ij}-\lambda K^{2})+\frac{\kappa^{2}\mu^{2}(\Lambda R-3\Lambda^{2})}{8(1-3
  \lambda)}\right]\\\\
  \mathcal{L}_{1}=\sqrt{g}N\left[\frac{\kappa^{2}\mu^{2}(1-4\lambda)}{32(1-3\lambda)}R^{2}-\frac{\kappa^{2}}{2\omega^{4}}(C_{ij}-\frac{\mu\omega^{2}}{2}R_{ij})(C^{ij}-\frac{\mu\omega^{2}}{2}R^{ij})\right]
  \\
\end{array}
\end{equation}

where, $\kappa^{2}$, $\lambda$, $\mu$, $\omega$ and $\Lambda$ are
constant parameters, and $C_{ij}$ is Cotton tensor (conserved and
traceless, vanishing for conformally flat metrics). The first two
terms in $\mathcal{L}_{0}$ are the kinetic terms, others in
$(\mathcal{L}_{0} +\mathcal{L}_{1})$ give the potential of the
theory in the so-called ``detailed-balance" form, and
$\mathcal{L}_{m}$ stands for the Lagrangian of other matter field.
Comparing the action to that of the general relativity, one can
see that the speed of light and the $cosmological$ Newton's
constant are \cite{Calcagni}, \cite{Jamil}, \cite{samarpita}
\begin{equation}
c=\frac{\kappa^{2}\mu}{4}\sqrt{\frac{\Lambda}{1-3\lambda}},~~~~~G_{c}=\frac{\kappa^{2}c}{16\pi(3\lambda-1)}
\end{equation}

It may be noted that when $\lambda=1$, $\mathcal{L}_{0}$ reduces
to the usual Lagrangian of Einstein's general relativity. Thus,
when $\lambda=1$, the general relativity is approximately
recovered at large distances.\\

We are considering the existence of both dark energy and dark
matter. Using the identifications in (10) one can rewrite the
field equations as \cite{samarpita}

\begin{equation}
H^{2} + \frac{k}{a^{2}}=\frac{8\pi
G_{c}}{3}\rho+\frac{k^{2}}{2\Lambda a^{4}}+\frac{\Lambda}{2}
\end{equation}
and
\begin{equation}
\dot{H}+\frac{3}{2}H^{2}+\frac{k}{2a^{2}}= -4 \pi
G_{c}p-\frac{k^{2}}{\Lambda a^{4}}+\frac{3\Lambda}{4}
\end{equation}

where, we have defined the Hubble parameter as
$H=\frac{\dot{a}}{a}$, $\Lambda$ is the cosmological constant. The
above equations (11) and (12) are derived in reference
\cite{samarpita}. The term proportional to $a^{-4}$ is the usual
``dark radiation term", present in HL cosmology. Also,
$\rho=\rho_{X}+\rho_{m}$ and $p=p_{X}+p_{m}$. Here, $X$ denotes
the dark energy component and $m$ denotes the dark matter
component. As there is no interaction, we have the conservation
equations

\begin{equation}
\dot{\rho}_{X}+3H(\rho_{X}+p_{X})=0~;~~~~\dot{\rho}_{m}+3H(\rho_{m}+p_{m})=0
\end{equation}
\\

\subsection{GRDE in Horava-Lifshitiz gravity}

From equations (5) and (6) we get the energy density of GRDE (we
assume $M_{pl}^{2}=1$). Thus,

\begin{equation}
\rho_{X}=3c^{2}\left[-6\left(2H^{2}+\dot{H}+\frac{k}{a^{2}}\right)(1-\eta)+H^{2}\eta\right]
\end{equation}

From the conservation equation (13)

\begin{equation}
\rho_{m}=\rho_{m0}a^{-3(1+w_{m})}
\end{equation}

Consequently,

\begin{equation}
H^{2}=\frac{8\pi
G_{c}}{3}\left[3c^{2}\left\{-6\left(2H^{2}+\dot{H}+\frac{k}{a^{2}}\right)(1-\eta)+H^{2}\eta\right\}+\rho_{m0}a^{-3(1+w_{m})}\right]+\frac{k^{2}}{2\Lambda
a^{4}}+\frac{\Lambda}{2}-\frac{k}{a^{2}}
\end{equation}

In (16), $H^{2}$ appears both in the left and the right hand side
and $\dot{H}$ appears in the right hand side. Solving the ordinary
differential equation (16) we can express $H^{2}$ as a function of
the scale factor $a$ as

\begin{equation}
H^{2}=\frac{1}{2}\left[\frac{\Lambda}{1+c^{2}(96-95\eta)}+\frac{1}{a^{4}}\left\{\frac{k^{2}}{\Lambda(1+c^{2}\eta)}+\frac{2a^{2}k(1-48c^{2}(-1+\eta))}{-1+c^{2}(48-47\eta)}+2a^{\frac{1+c^{2}\eta}{24c^{2}(-1+\eta)}}C_{1}\right\}\right]-\frac{\rho_{m0}a^{-3(1+w_{m})}}{3(-1+c^{2}(-96+95\eta))}
\end{equation}

Dividing both sides of the above Friedman equation by $3H_{0}^{2}$
we get

\begin{equation}
\mathcal{H}^{2}=\frac{(1-48c^{2}(-1+\eta))\Omega_{k0}}{a^{2}(-1+c^{2}(-48+47\eta))}+\frac{\Omega_{k0}^{2}}{2a^{4}(1+c^{2}\eta)\Omega_{\Lambda
0}}+\frac{\Omega_{\Lambda
0}}{2(1+c^{2}(96-95\eta))}+\frac{a^{-3(1+w_{m})}\Omega_{m0}}{3(-1+c^{2}(-96+95\eta))}+a^{-4+\frac{1+c^{2}\eta}{24c^{2}(-1+\eta)}}f_{0}
\end{equation}

where, the relative densities are

\begin{equation}
\Omega_{\Lambda
0}=\frac{\Lambda}{3H_{0}^{2}}~;~~\Omega_{k0}=\frac{k}{3H_{0}^{2}}~;~~\Omega_{m0}=\frac{\rho_{m}}{3H_{0}^{2}}
\end{equation}

and $f_{0}=\frac{C_{1}}{3H_{0}^{2}}$ is the integration constant.
Using the above expression of $\mathcal{H}^{2}$ in (14) after
dividing (14) by $3H_{0}^{2}$ we get

\begin{equation}
\begin{array}{c}
 \Omega_{X0}=3c^{2}\left[\frac{a^{\frac{1}{24}\left(-95+\frac{1+c^{2}}{c^{2}(-1+\eta)}\right)}f_{0}(1+9c^{2}\eta)}{8c^{2}}+\frac{(1-54c^{2}(-1+\eta))\eta\Omega_{k0}}{a^{2}(-1+c^{2}(-48+47\eta))}+\frac{\eta
\Omega_{k0}^{2}}{2a^{4}\Omega_{\Lambda0}(1+c^{2}\eta)}\right.\\
 \left.-\frac{a^{-3(1+w_{m})}(-12+13\eta)\left(3a^{3(1+w_{m})}\Omega_{\Lambda0}-2\Omega_{m0}\right)}{6\left(-1+c^{2}(-96+95\eta)\right)}\right] \\
\end{array}
\end{equation}
\\\\
\begin{figure}
\includegraphics[height=2.1in]{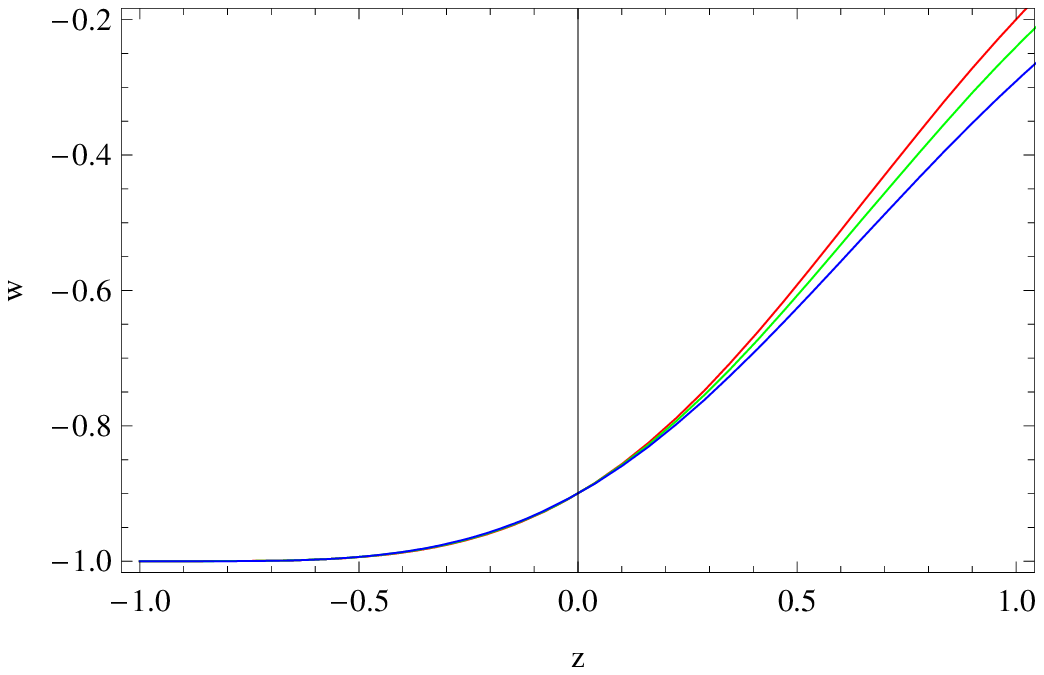}~~~\\
\vspace{1mm} ~~~~~~~~~~~~Fig.1\\
\vspace{6mm} Fig. 1 shows the variation of the equation of state
parameter  $w$ against $z$ in for GRDE in the Horava-Lifshitz
gravity. The red, green and blue lines correspond to
$c^{2}=0.2,~0.5$ and $-0.8$ respectively. Also, we have taken
$\Omega_{mo}=0.27,~~\Omega_{k0}=5\times10^{-5},~~\Omega_{\Lambda0}=1\times10^{-6}$.

\vspace{6mm}

\end{figure}
\begin{figure}
\includegraphics[height=2.1in]{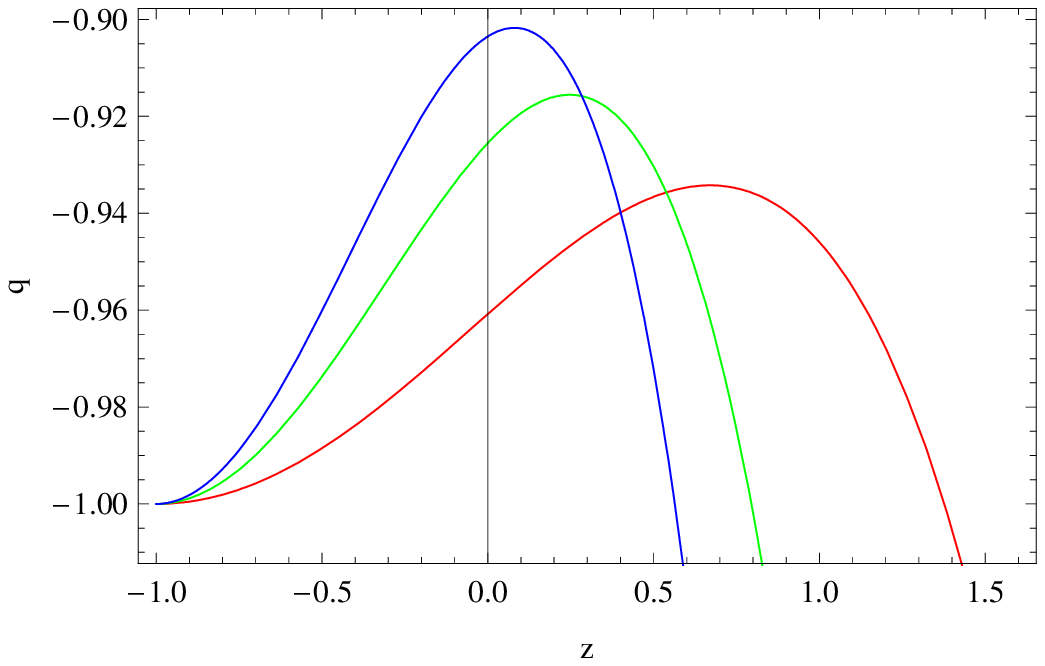}~~~\\
\vspace{1mm} ~~~~~~~~~~~~Fig.2\\
\vspace{6mm} Fig. 2 shows the variation of the deceleration
parameter $q$ against $z$ in for GRDE in the Horava-Lifshitz
gravity. The red, green and blue lines correspond to
$c^{2}=0.2,~0.5$ and $-0.8$ respectively. Also, we have taken
$\Omega_{mo}=0.27,~~\Omega_{k0}=5\times10^{-5},~~\Omega_{\Lambda0}=1\times10^{-6}$.

\vspace{6mm}

\end{figure}

\begin{figure}
\includegraphics[height=2.8in]{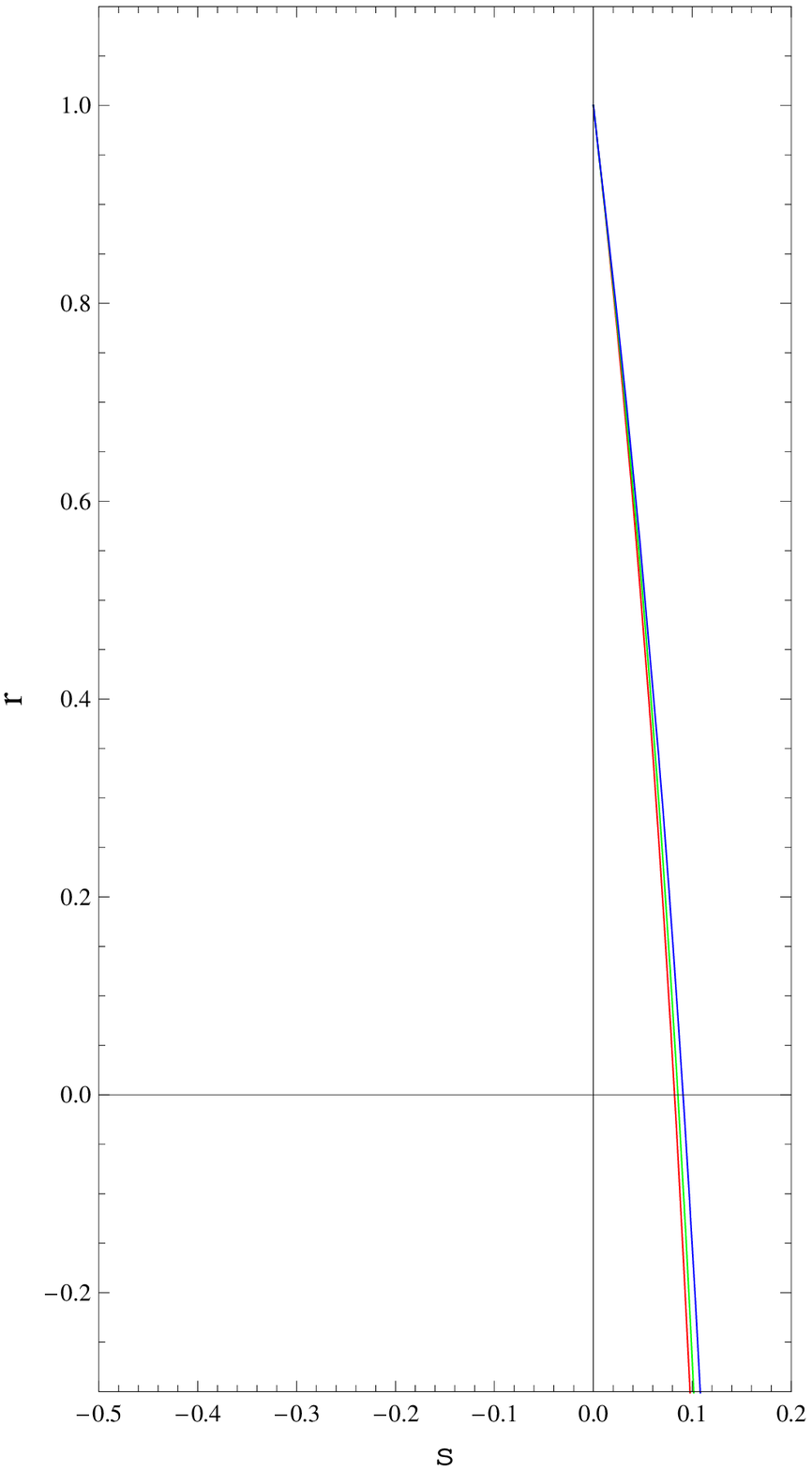}~~~\\
\vspace{1mm} ~~~~~~~~~~~~Fig.3\\
\vspace{6mm} Fig. 3 shows $r-s$ trajectory for GRDE in the
Horava-Lifshitz gravity. The red, green and blue lines correspond
to $c^{2}=0.2,~0.5$ and $-0.8$ respectively.

\vspace{6mm}

\end{figure}
Using (18) and (20) in the conservation equation and considering
$w_{0}$ as the present value of the EoS parameter i.e.
$w_{0}=p_{X}/\rho_{X}$ ($a=1$), the expression for $f_{0}$ can be
determined as

\begin{equation}
f_{0}=-\frac{192c^{4}(-A_{1}+3A_{2}(1+3w_{0}))(1-\eta)}{(1+9c^{2}\eta)(1+24c^{2}(1-33w_{0})-c^{2}\eta
(23-72w_{0}))}
\end{equation}

where,

\begin{equation}
A_{1}=\frac{2(-1+54c^{2}(-1+\eta))\eta
\Omega_{k0}}{-1+c^{2}(-48+47\eta)}-\frac{2\eta\Omega_{k0}^{2}}{(1+c^{2}\eta)\Omega_{\Lambda0}}
\end{equation}

\begin{equation}
A_{2}=\frac{(-1+54c^{2}(-1+\eta))\eta
\Omega_{k0}}{-1+c^{2}(-48+47\eta)}-\frac{\eta\Omega_{k0}^{2}}{2(1+c^{2}\eta)\Omega_{\Lambda0}}-\frac{a^{-3(1+w_{m})}(12-13\eta)(3a^{3(1+w_{m})}\Omega_{\Lambda0}-2\Omega_{m0})}{6(-1+c^{2}(-96+95\eta))}
\end{equation}

Now we consider the deceleration parameter $q=$ for the same
model. The deceleration parameter is given by \cite{Varun}
\begin{equation}
q=-1-\frac{\dot{H}}{H^{2}}=-1-\frac{a}{2\mathcal{H}^{2}}\frac{d\mathcal{H}^{2}}{da}
\end{equation}

Using the results obtained above we get the form of $q$ as

\begin{equation}
q=-1-\frac{a^{-2+\frac{1+c^{2}\eta}{48c^{2}(-1+\eta)}}f_{0}\left(-4+\frac{1+c^{2}\eta}{24c^{2}(-1+\eta)}\right)-\frac{(1-48c^{2}(-1+\eta))\Omega_{k0}}{a^{2}(-1+c^{2}(-48+47\eta))}-\frac{\Omega_{k0}^{2}}{a^{4}\Omega_{\Lambda0}(1+c^{2}\eta)}}
{a^{-5+\frac{1+c^{2}\eta}{24c^{2}(-1+\eta)}}f_{0}+\frac{(1-48c^{2}(-1+\eta))\Omega_{k0}}{a^{2}(-1+c^{2}(-48+47\eta))}+\frac{\Omega_{k0}^{2}}{2a^{4}\Omega_{\Lambda0}(1+c^{2}\eta)}+\frac{\Omega_{\Lambda0}}{2(1+c^{2}(96-95\eta))}+\frac{a^{-3(1+w_{m})}\Omega_{m0}}{3(-1+c^{2}(-96+95\eta))}}
\end{equation}
\\
Now we discuss the statefinder diagnostic pair i.e., $\{r,s\}$
parameters that are of the following form \cite{chattopadhyaytach}

\begin{equation}
r=\frac{\dddot{a}}{a\mathcal{H}^{3}}=1+3\frac{\dot{\mathcal{H}}}{\mathcal{H}^{2}}+\frac{\ddot{\mathcal{H}}}{\mathcal{H}^{3}}
\end{equation}
and
\begin{equation}
s=\frac{r-1}{3\left(q-\frac{1}{2}\right)}=-\frac{3\mathcal{H}\dot{\mathcal{H}}+\ddot{\mathcal{H}}}{3\mathcal{H}(2\dot{\mathcal{H}}+3\mathcal{H}^{2})}
\end{equation}

The $\{r,s\}$ pair is now constructed based on $\mathcal{H}$
expressed earlier. The $r-s$ plane would be constructed and an
$r-s$ trajectory would be created and would be discussed in the
next subsection.
\\
\subsection{Discussion}

In this work we have considered generalized Ricci dark energy in
the Horava-Lifshitz gravity. We have reconstructed the Hubble's
parameter in terms of fractional densities. We have viewed the
equation of state parameter in this situation. In figure 1 we have
plotted the equation of state parameter $w$ for generalized Ricci
dark energy taking $c^{2}=0.2,~0.5$ and $-0.8$. We have taken
$\eta=\frac{1}{13}$ and $w_{0}=-0.9$. For different values of
$c^{2}$ we plot $w$. For the current model we find that the
equation of state parameter stays above $-1$ for all values of
$c^{2}$. This indicates quintessence like behavior.  The
deceleration parameter expressed above is presented in figure 2,
which shows that $q$ is staying in the negative side for all
redshifts in the case of GRDE considered in Horava-Lifshitz
gravity. This indicates an ever accelerating universe. The $r-s$
trajectory is presented in figure 3. Here we find that for all
values of $c^{2}$ the $r-s$ trajectory is confined within the
first and fourth quadrant of the $r-s$ plane. The trajectory
begins from $\{r=1,s=0\}$ and then $r\rightarrow-\infty$ for
finite $s$. Thus we see that the trajectory passes through and not
go beyond the $\Lambda$CDM phase of the universe when we consider
generalized Ricci dark energy in
Horava-Lifshitz gravity.\\
The present study deviates from the earlier studies in the
following aspects: Contrary to earlier works \cite{Xu1},
\cite{FengLi} done on Einstein gravity the present study is made
in Horava-Lifshitz gravity, which is a modified gravity theory. In
\cite{Feng}, Ricci dark energy was considered in $f(R)$ gravity.
In the present paper we have considered generalized Ricci dark
energy in Horava-Lifshitz gravity. In reference
\cite{chattopadhyaychameleon}, Ricci dark energy was considered on
chameleon field with the choice of scale factor in the form of
emergent universe. However, in the present paper we have not
considered any particular choice of scale factor. A study of
statefinder diagnostics of Ricci dark energy was done by
\cite{fengstatefinder}, where it was found that the $\{r-s\}$
trajectory is a vertical segment, i.e. $s$ is a constant during
the evolution of the universe for a particular choice of $c^{2}$.
However, when we consider generalized Ricci dark energy in
Horava-Lifshitz gravity, we find that the trajectory begins from
$\{r=1,s=0\}$ and then $r\rightarrow-\infty$. In the work by
\cite{Zhang7}, $c^{2}$ was taken as 0.3, 0.4, 0.5, and 0.6 and it
was observed that for $c^{2}<1/2$ the equation of state parameter
$w<-1$, i.e. behaves like quintom and for $c^{2}>1/2$ the equation
of state parameter $w>-1$, i.e. behaves like a quintessence.
However, the present study shows that when generalized Ricci dark
energy is considered in Horava-Lifshitz gravity, the equation of
state parameter always exhibits quintessence-like bahavior
irrespective of $c^{2}>,~=,~<1/2$.
\\\\
{\bf Acknowledgement:}\\
The author is thankful to Dr. Ujjal Debnath for some valuable
discussions related to this work. The author wishes to sincerely
acknowledge the warm hospitality provided by Inter-University
Centre for Astronomy and Astrophysics (IUCAA), Pune, India, where
part of the work was carried out during a scientific visit in
January, 2011.
\\
\\\\

\end{document}